\documentclass[conference, 9pt]{IEEEtran}
\usepackage{amsmath}
\IEEEoverridecommandlockouts
\usepackage{cite}
\usepackage{amsmath,amssymb,amsfonts}
\usepackage{algorithm}
\usepackage[noend]{algpseudocode}
\usepackage{graphicx}
\usepackage{textcomp}
\usepackage{xcolor}
\usepackage{acronym}
\usepackage{multirow}
\usepackage{amssymb}
\usepackage{pifont}
\usepackage{todonotes}
\usepackage{graphicx}
\usepackage{tabularx}
\graphicspath{{figs/}}
\usepackage{xcolor}
\usepackage[hidelinks, colorlinks=true, linkcolor=blue, filecolor=blue, citecolor=blue, urlcolor=blue]{hyperref}
\usepackage{rotating}
\usepackage{booktabs}
\usepackage[absolute,overlay]{textpos}

\newcommand{\orcidicon}[1]{\href{https://orcid.org/#1}{\includegraphics[width=10pt]{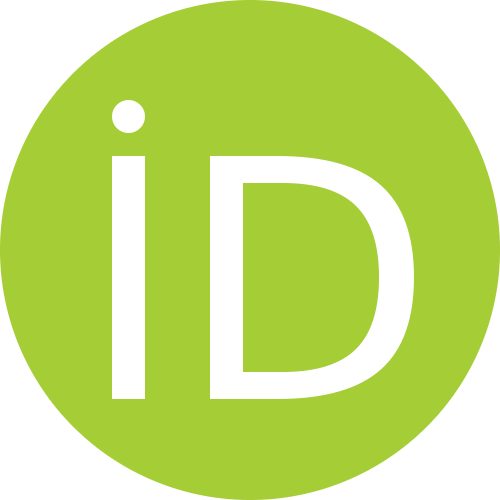}}}

\algrenewcommand\algorithmicthen{}

\newcommand{\cmark}{\ding{51}} 
\newcommand{\xmark}{\ding{55}} 

\def\BibTeX{{\rm B\kern-.05em{\sc i\kern-.025em b}\kern-.08em
    T\kern-.1667em\lower.7ex\hbox{E}\kern-.125emX}}

\definecolor{ceil}{rgb}{0.57, 0.63, 0.81}

\begin{document}

\acrodef{api}[API]{Application Programming Interface}
\acrodef{ai}[AI]{Artificial Intelligence}
\acrodef{dla}[DLA]{Deep Learning Accelerator}
\acrodef{cpu}[CPU]{CPU}
\acrodef{gpu}[GPU]{Graphics Processing Unit}
\acrodef{npu}[NPU]{Neural Processing Unit}
\acrodef{tpu}[TPU]{Tensor Processing Unit}
\acrodef{xpu}[XPU]{Custom Acceleration Processing Unit}
\acrodef{dvfs}[DVFS]{Dynamic Voltage/Frequency Scaling}
\acrodef{hmp}[HMP]{Heterogeneous Multi-Processing}
\acrodef{hsa}[HSA]{Heterogeneous System Architecture}
\acrodef{os}[OS]{Operating System}
\acrodef{tdp}[TDP]{Thermal Design Power}
\acrodef{gn}[GN]{Gateway Node}
\acrodef{ln}[LN]{Local Node}
\acrodef{iot}[IoT]{Internet of Things}
\acrodef{ml}[ML]{Machine Learning}
\acrodef{qos}[QoS]{Quality of Service}
\acrodef{ftp}[FTP]{Fused Tile Partitioning}
\acrodef{aofl}[AOFL]{Adaptive Optimal Fused-layer}
\acrodef{socs}[SoCs] {System-on-Chips}
\acrodef{rtm}[RTM]{Run-time Resource Management}
\acrodef{dnn}[DNN]{Deep Neural Network}
\acrodef{dop}[DoP]{Degree of Parallelism}
\acrodef{ppw}[PPW]{Performance per Watt}
\acrodef{os}[OS]{Operating System}
\acrodef{fsm}[FSM]{Finite State Machine}
\acrodef{lan}[LAN]{Local Area Network}
\acrodef{wlan}[WLAN]{Wireless Local Area Network}
\acrodef{elan}[ELAN]{Ethernet Local Area Network}
\acrodef{ilp}[ILP]{Integer Linear Programming}
\acrodef{eeg}[EEG]{Electroenceophelogram}
\acrodef{rl}[RL]{Reinforcement Learning}
\acrodef{bodp}[BODP]{Biased One-Dimensional Partition}
\acrodef{dl}[DL]{Deep Learning}
\acrodef{morl}[MORL]{Multi-Objective Reinforcement Learning}
\acrodef{momdp}[MOMDP]{Multi-Objective Markov Decision Process}
\acrodef{mdp}[MDP]{Markov's Decision Process}
\acrodef{Sarsa}[Sarsa]{State-Action-Reward-State’-Action’}
\acrodef{onnx}[ONNX]{Open Neural Network Exchange}
\acrodef{ar}[AR]{Augmented Reality}
\acrodef{vr}[VR]{Virtual Reality}
\acrodef{mr}[MR]{Mixed Reality}
\acrodef{dag}[DAG]{Directed Acyclic Graphs}
\acrodef{mac}[MAC]{Multiply–Accumulate operation}
\acrodef{dse}[DSE]{Design Space Exploration}
\acrodef{ppo}[PPO]{Proximal Policy Optimization}
\acrodef{soa}[SoA]{State-of-the-Art}
\acrodef{cnn}[CNN]{Convolution Neural Network}
\acrodef{fcfs}[FCSFS]{First-come-first-serve}
\acrodef{mas}[MAS]{Multi-Agent Systems}
\acrodef{cai}[cAI]{Compound Artificial Intelligence}
\acrodef{llm}[LLM]{Large Language Model}
\acrodef{lm}[LM]{Language Model}
\acrodef{mri}[MRI]{Magnetic Resonance Imaging}
\acrodef{fps}[FPS]{Frames Per Second}
\acrodef{tm}[TM]{Task Migration}
\acrodef{cs}[CS]{Context Switching}
\acrodef{sm}[SM]{Streaming Multiprocessor}
\acrodef{vit}[ViT]{Vision Transformer}

\title{Twill: Scheduling Compound AI Systems on Heterogeneous Mobile Edge Platforms \\
}

\author{
\IEEEauthorblockN{Zain Taufique \href{https://orcid.org/0000-0002-5298-6049}{\includegraphics[width=8pt]{orcid.png}}}  
\IEEEauthorblockA{
\textit{University of Turku}\\
Turku, Finland \\
zatauf@utu.fi}
\and
\IEEEauthorblockN{Aman Vyas \href{https://orcid.org/0009-0002-8514-4582}{\includegraphics[width=8pt]{orcid.png}}}  
\IEEEauthorblockA{
\textit{University of Turku}\\
Turku, Finland \\
amvyas@utu.fi}
\and
\IEEEauthorblockN{Antonio Miele \href{https://orcid.org/0000-0003-3197-0723}{\includegraphics[width=8pt]{orcid.png}}}  
\IEEEauthorblockA{
\textit{Politecnico di Milano}\\
Milan, Italy \\
antonio.miele@polimi.it}
\and
\IEEEauthorblockN{Pasi Liljeberg \href{https://orcid.org/0000-0002-9392-3589}{\includegraphics[width=8pt]{orcid.png}}}  
\IEEEauthorblockA{
\textit{University of Turku}\\
Turku, Finland \\
pasi.liljeberg@utu.fi}
\and
\IEEEauthorblockN{Anil Kanduri \href{https://orcid.org/0000-0003-3188-8703}{\includegraphics[width=8pt]{orcid.png}}}  
\IEEEauthorblockA{
\textit{University of Turku}\\
Turku, Finland \\
spakan@utu.fi}
}

\maketitle

\begin{textblock*}{3cm}(20cm,8cm) 
\rotatebox{90}{\textit{Accepted in International Conference on Computer-Aided Design (ICCAD) 2025}}
\end{textblock*}

\begin{abstract}
Compound AI (cAI) systems chain multiple AI models to solve complex problems. cAI systems are typically composed of deep neural networks (DNNs), transformers, and large language models (LLMs), exhibiting a high degree of computational diversity and dynamic workload variation. Deploying cAI services on mobile edge platforms poses a significant challenge in scheduling concurrent DNN-transformer inference tasks, which arrive dynamically in an unknown sequence. Existing mobile edge AI inference strategies manage multi-DNN or transformer-only workloads, relying on design-time profiling, and cannot handle concurrent inference of DNNs and transformers required by cAI systems. In this work, we address the challenge of scheduling cAI systems on heterogeneous mobile edge platforms. We present \textit{Twill}, a run-time framework to handle concurrent inference requests of cAI workloads through task affinity-aware cluster mapping and migration, priority-aware task freezing/unfreezing, and \ac{dvfs}, while minimizing inference latency within power budgets. We implement and deploy our \textit{Twill} framework on the Nvidia Jetson Orin NX platform. We evaluate \textit{Twill} against state-of-the-art edge AI inference techniques over contemporary DNNs and LLMs, reducing inference latency by 54\% on average, while honoring power budgets. 

\end{abstract}


\acresetall
\acused{gpu}
\acused{cpu}
\acused{ai}

\section{Introduction}
\begin{figure}
    \centering
    \vspace{-3pt}
    \includegraphics[width=0.99\linewidth]{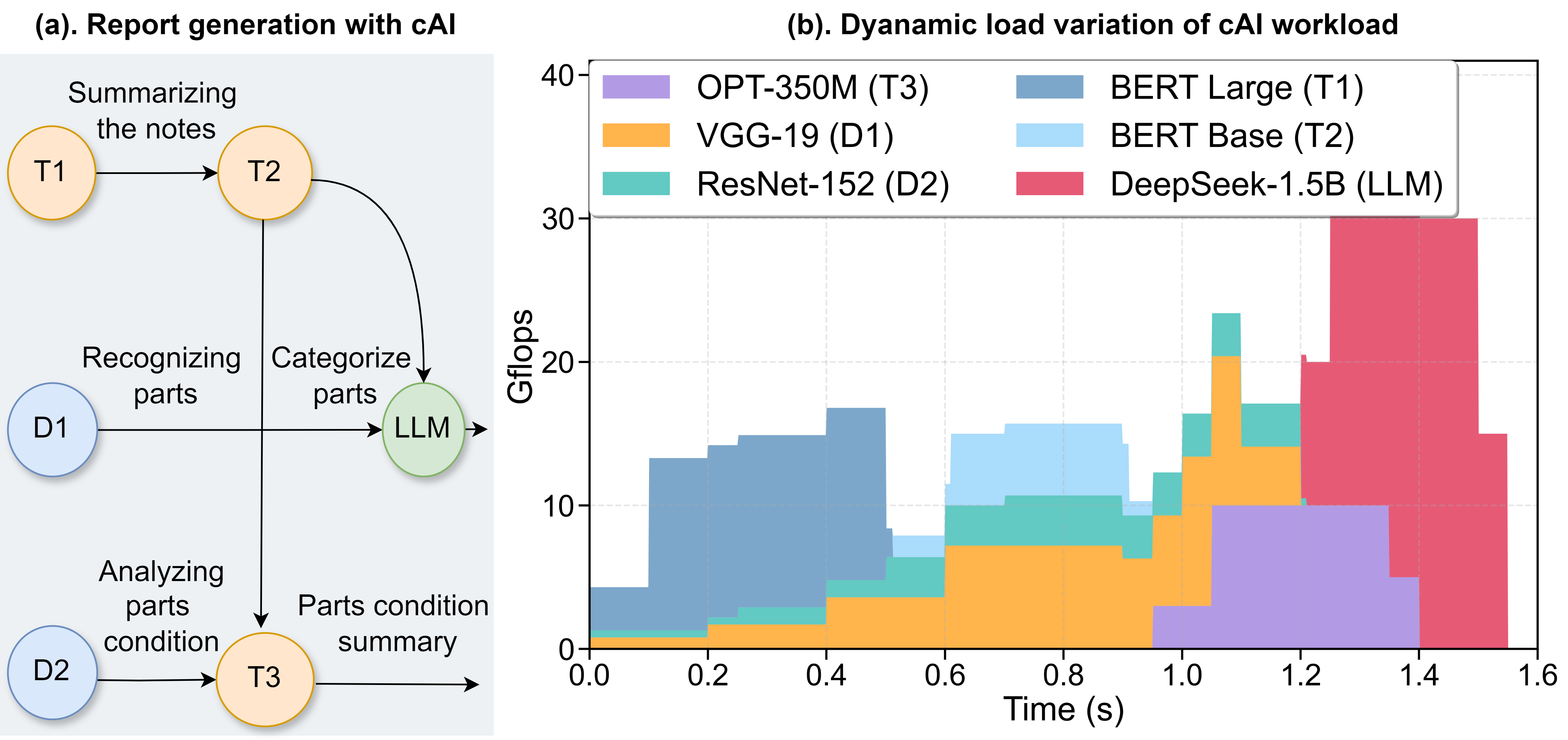}
    \vspace{-9pt}
    \caption{Exemplar \ac{cai} system. (a) Task graph for \ac{cai} system chaining multiple models, (b) Run-time workload variation and compute diversity.}
    \label{fig.mot-1}
    \vspace{-15pt}
\end{figure}

\ac{ai} applications are rapidly evolving from monolithic models towards \ac{cai} systems, which integrate multiple task-specific models and components to solve complex problems~\cite{compound_ai,chen2024more,chen2025optimizing}. Emerging \ac{cai} systems combine \acp{llm} with \acp{dnn} for providing novel services such as conversational language agents~\cite{chen2025optimizing,valmeekam2023can,chen2024more,wu2023autogen}, augmented and virtual reality (AR/VR) gear, and interactive autonomous vehicles~\cite{cui2024drive}. \ac{cai} systems offer compositional flexibility by selectively chaining multiple transformer (both encoder and generative) and \ac{dnn} models at run-time~\cite{langchain,langbase}. Figure~\ref{fig.mot-1}(a) shows a conceptual example of a \ac{cai} workload designed as a task graph for generating a maintenance report from the input images and text given by the user. In this example, \ac{dnn} models (\texttt{D1}: \texttt{VGG-19} and \texttt{D2}: \texttt{ResNet-152}) are used for image classification, and object detection, transformer models (\texttt{T1}: \texttt{Bert-base} and \texttt{T2}: \texttt{Bert-large}) are used for text summarizing and classification, and generative transformers (\texttt{T3}: \texttt{OPT-350M} and \texttt{LLM}: \texttt{Deepseek-R1}) are used for reasoning and report generation. Each model is responsible for extracting key features from the given input and sending the output to the subsequent models to perform collaborative tasks. \texttt{T1}, \texttt{D1}, and \texttt{D2} are exclusive inference tasks that can run simultaneously, while \texttt{T2}, \texttt{T3}, and LLM are dependent on the outputs of other models. We deployed the exemplar \ac{cai} system on the Nvidia Jetson Orin NX platform. Figure~\ref{fig.mot-1}(b) shows performance demands (in GFlops)  of the exemplar \ac{cai} system. In this example, \ac{cai} system requires concurrent execution of (i) multiple \ac{dnn} and a transformer ($t = 0s\text{ - }0.9s$), (ii) multiple \ac{dnn} and a generative transformer ($t =  0.9s \text{ - } 1.2s$), and (iii) multiple generative transformers ($t=1.2s \text{ - }1.5s$). This demonstrates a high degree of computational diversity and dynamic workload variation with \ac{cai} systems. Thus, the primary requirement for implementing \ac{cai} systems is concurrent execution of transformers and \ac{dnn} models, where inference requests are computationally diverse and variable at run-time based on user requirements~\cite{langchain,langbase}. On the other hand, there is an increasing demand to deploy \ac{cai} inference services on user-end mobile and edge platforms to address the latency, bandwidth, and privacy challenges of the cloud infrastructure~\cite{AutoScale}. However, running \ac{cai} workloads on heterogeneous mobile edge platforms poses significant challenges in scheduling concurrent execution of transformers and \acp{dnn} while reducing inference latency within the power constraints~\cite{band}. 

Despite being extremely resource-constrained, mobile edge platforms support AI inference services through powerful mobile GPUs and domain-specific \acp{dla}~\cite{AutoScale,AutoDice,band}. For example, smart phones \cite{apple_intelligence_2024}, smart glasses (e.g. Ray-ban Meta~\cite{rayban}) and AR gear (e.g. Apple Vision Pro~\cite{vision_pro}) etc., integrate \acp{gpu} and \acp{dla}, such as \acp{npu} or \acp{tpu}~\cite{qualcommxr}. Existing edge AI orchestration techniques have optimized scheduling of multi-DNN workloads~\cite{AutoDice,tango,omniboost,moc,pipeit} and transformer models~\cite{easter,pipebert} on mobile platforms with GPUs and \acp{dla}. Some of these techniques do not consider transformer models and/or \acp{dla}~\cite{tango,pipeit,moc,omniboost,HaX_CoNN}. Using these techniques for \ac{cai} systems results in significantly higher latency and power consumption, suffering from shared resource contention by mapping both \ac{dnn} and transformer models on the GPU cluster. Techniques that consider transformer models and \acp{dla}~\cite{pipebert,band,mapformer} ignore the lack of operation-level support for transformers on \acp{dla}. Existing \acp{dla} are optimized specifically for \acp{dnn} and do not support the majority of key operations required for transformer inference\cite{rockchip_NPU, dla_manual}. Executing a transformer inference on \acp{dla} leads to frequent GPU \textit{fallbacks}, i.e., moving the execution of operations that are unsupported on the \ac{dla} to the GPU. This causes heavy memory access overhead and results in significantly higher inference latency. 
With a lack of operator support on \ac{dla}, existing techniques execute transformer models inevitably on the \ac{gpu}~\cite{mapformer,pipebert}. Running transformers on the GPU is a fair choice, as long as a standalone transformer model inference is required. However, \ac{cai} workloads require concurrent execution of multiple \acp{dnn} and transformers. A trivial solution is to map \acp{dnn} on \ac{dla} and transformers on GPU; however, this approach is suitable only when the order of inference requests is known at design time. For example, consider a scenario where a transformer inference request arrives when a \ac{dnn} is already running on the GPU. In this case, the transformer inference request is either queued until the \ac{dnn} inference task is completed, or both \ac{dnn} and transformer run concurrently on the GPU, suffering shared resource contention. With no consideration on operator-level support and the need for concurrent \ac{dnn}-transformer execution, existing edge AI inference techniques are not adaptable for \ac{cai} workloads. 


Orchestrating \ac{cai} workloads requires a run-time adaptive scheduling strategy that considers dynamic inference requests and migrates diverse inference tasks among GPU and \ac{dla} clusters to minimize overall inference latency.
In this work, we address the challenge of scheduling \ac{cai} workloads through model affinity and priority-aware task-to-cluster mapping and migration. We present \textit{Twill}, a novel framework for run-time partitioning, distribution, and scheduling of \ac{cai} workloads for heterogeneous mobile edge platforms. We analyze each inference request to determine the affinity of a model towards a cluster and schedule concurrent inference requests on feasible clusters to minimize latency. Our approach adaptively migrates inference requests among feasible clusters and/or freezes inference requests based on priority to accommodate other concurrently running inference tasks. Further, we monitor the run-time power consumption and actuate \ac{dvfs} to honor \ac{tdp} constraint. We implemented and deployed the \textit{Twill} framework on the Jetson Orin NX embedded platform and evaluated over contemporary \ac{cai} workloads comprising of widely used \acp{dnn} (\texttt{VGG-19, ResNet-152}, and \texttt{EfficientNet-b4}), transformer (\texttt{Bert-base, Bert-large, ViT-base}, and \texttt{ViT-large}) models, and \acp{llm} (\texttt{Deepseek-R1, Gemma-3}). Our novel contributions are:
\begin{itemize}
    \item \textit{Twill}, a run-time framework for partitioning, distribution, and scheduling of \ac{cai} workloads on heterogeneous mobile edge platforms.
    \item Online heuristic model for profiling \ac{dnn}, transformer, and \ac{llm} inference requests to determine affinity of the inference task towards a compute cluster (\ac{gpu}/\ac{dla}).
    \item Inference task-to-cluster mapping and migration, and priority-aware task freezing for concurrent execution of \ac{dnn}, transformer, and \ac{llm} inference requests, and \ac{dvfs} actuation for power capping.
    \item Evaluation of \textit{Twill} on commercial edge \ac{ai} platform Jetson Orin NX using contemporary \ac{cai} workloads.
\end{itemize}

\noindent \textit{Manuscript Organization:} Section~\ref{sec.mot} provides background and motivation for our proposed approach, Section~\ref{sec.arch} presents an overview of our run-time management framework infrastructure, Section~\ref{sec.exp} presents an evaluation of our proposed solution against other relevant strategies, followed by conclusions in Section~\ref{sec.conc}.

\section{Background and Motivation}\label{sec.mot}
\subsection{Impact of scheduling on cAI workloads}
\begin{figure}
    \centering
    \vspace{-3pt}
    \includegraphics[width=0.99\linewidth]{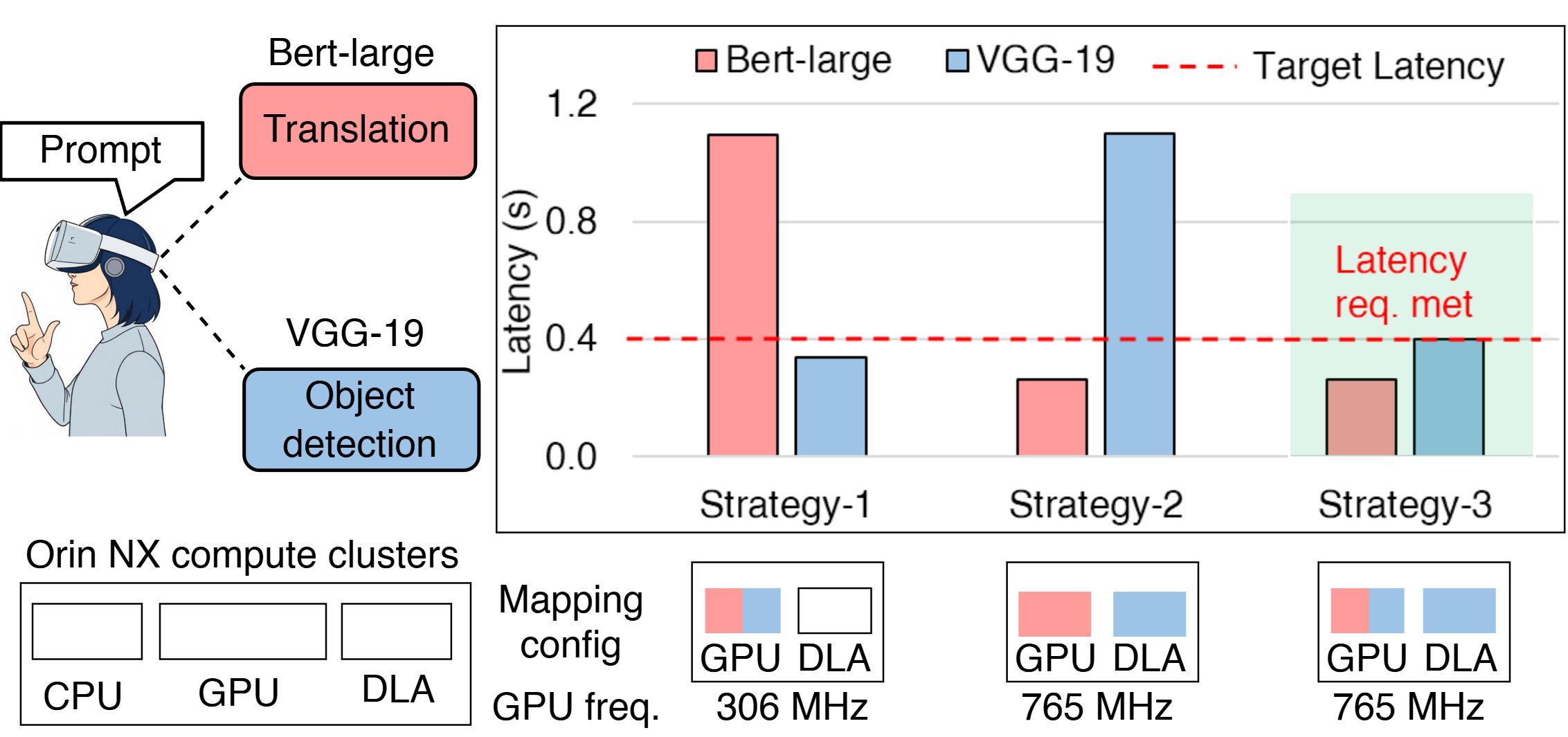}
    \vspace{-9pt}
    \caption{Inference latency of \ac{cai} system across different execution strategies}
    \vspace{-15pt}
    \label{fig.intro}
\end{figure}

Figure~\ref{fig.intro} presents a realistic \ac{cai} workload example of a multi-linguistic translation application on a \ac{vr} gear platform that uses the \texttt{VGG-19}~\cite{vgg} model to detect the text and \texttt{Bert-large} transformer~\cite{bert} for translation to a required language. 
For simplicity of demonstration, we chose \ac{cai} workload with two models that have no dependencies; in general, \ac{cai} workloads can be composed of several diverse models with data dependencies. In this example, the user generates run-time inference requests based on his/her activity, and the arrival time of these requests is unknown to the baseline platforms at design time. In Figure~\ref{fig.intro}, we compare three workload scheduling strategies for inferring \texttt{VGG-19} and \texttt{Bert-large} on the Nvidia Jetson Orin NX platform~\cite{orin_nx} that includes \ac{cpu}, \ac{gpu}, and \ac{dla} clusters. In this example, we set the batch size of \texttt{VGG-19} model to 32 images, the input sequence length of \texttt{Bert-large} to 128 tokens, and the minimum inference latency to 400ms, representing a real-world scenario. \textit{Strategy-1} is representative of existing multi-DNN scheduling strategies \cite{tango, moc, omniboost} that map both applications on \ac{gpu}. Here, \texttt{Bert-large} has to wait for the execution of \texttt{VGG-19} to use the GPU resources exceeding the minimum latency requirement. \textit{Strategy-2} maps \texttt{VGG-19} to \ac{dla} and \texttt{Bert-large} to \ac{gpu} while increasing the \ac{gpu} frequency. Here, \texttt{Bert-large} meets the latency requirements while \texttt{VGG-19} is lagging due to the low performance of \ac{dla}. \textit{Strategy-2} is representative of state-of-the-art edge AI inference techniques that handle transformer models~\cite{mapformer, HaX_CoNN, band}. Finally, \textit{Strategy-3} (representative of proposed \textit{Twill} approach) successfully meets the latency requirements by (i) partitioning and mapping \texttt{VGG-19} on \ac{dla}, (ii) mapping \texttt{Bert-large} on \ac{gpu}, and (iii) re-allocating the \ac{gpu} resources to \texttt{VGG-19} once \texttt{Bert-large} finishes execution. It should be noted that the scheduling strategies primarily map the inference requests to \ac{gpu} and \ac{dla} clusters, since \acp{cpu} have extremely high latency for such computationally-intensive workloads.


\subsection{Dynamics of scheduling DNN-Transformer concurrently}
Scheduling \ac{dnn} and transformer models concurrently requires joint actuation of cluster migration, run-time task freezing/unfreezing, and \textit{\ac{dvfs}}. We demonstrate the efficacy of each knob actuation setting under different workload scenarios and scheduling strategies. We use \acp{dnn} \texttt{ResNet-152} and \texttt{VGG-19}, and transformer \texttt{Bert-base}, which are run on the Jetson Orin NX platform. 

\begin{figure}
    \centering
    \vspace{-3pt}
    \includegraphics[width=0.99\linewidth]{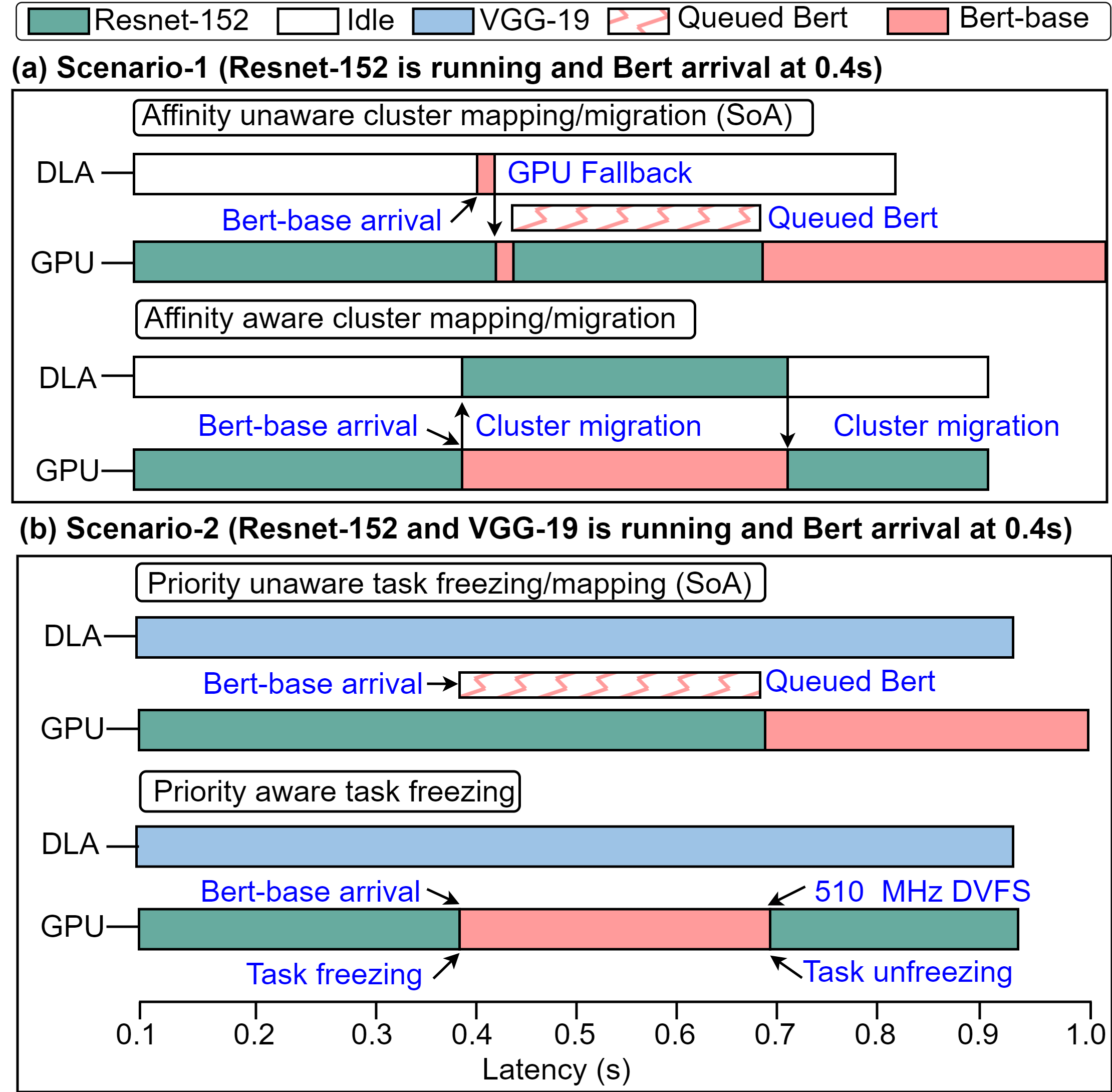}
    \vspace{-12pt}
    \caption{Various knob actuation scenarios, highlighting lack of priority and affinity adjustments at runtime in SoA strategies.}
    \vspace{-15pt}
    \label{fig.mot-2}
\end{figure}

\noindent \textbf{Scenario-1.} Figure~\ref{fig.mot-2}(a) shows workload scenario where a \ac{dnn} model \texttt{ResNet-152} is running on \ac{gpu} and inference request for transformer \texttt{Bert-base} arrives at $t=0.4s$. The SoA approaches~\cite{mapformer, omniboost, HaX_CoNN, band} lack affinity awareness, and map the incoming \texttt{Bert-base} on the available \ac{dla} cluster. However, \texttt{Bert-base} falls back to the GPU due to a lack of operator support. At this point, if GPU memory is insufficient to run both \texttt{ResNet-152} and \texttt{Bert-base}, \texttt{Bert-base} is queued until \texttt{ResNet-152} finishes execution. This results in higher inference latency of \texttt{Bert-base}. If sufficient GPU memory is available, both inference tasks will run on the GPU; however, this still leads to higher inference latency of both tasks due to shared resource contention. In contrast, our proposed strategy considers the affinity of \texttt{Bert-base} towards GPU upon on arrival of the inference request. Consequently, (i)~\texttt{ResNet-152} is migrated to \ac{dla} to accommodate the transformer model on GPU, (ii)~\texttt{Bert-base} is mapped onto the GPU, (iii)~\texttt{ResNet-152} is migrated to GPU once \texttt{Bert-base} finishes the execution at $t=0.71s$, freeing up the GPU cluster. This approach minimizes the overall inference latency and waiting time in the execution queue, while maximizing resource utilization. 
\newline\noindent \textbf{Scenario-2.} Figure~\ref{fig.mot-2}(b) shows workload scenario where two \ac{dnn} models viz., \texttt{ResNet-152} on \ac{gpu} and \texttt{VGG-19} on \ac{dla} are concurrently running, and inference request for transformer \texttt{Bert-base} arrives at $t=0.4s$. In this scenario, both GPU and \ac{dla} clusters are busy, and memory utilization is relatively higher. Under this workload, SoA approaches~\cite{moc, mapformer,HaX_CoNN}, without dynamic mapping/migration capabilities, queue \texttt{Bert-base} until \texttt{Resnet-152} completes execution on GPU. Inevitably, this leads to higher waiting time in the execution queue and inference latency of \texttt{Bert-base}. In this case, our proposed strategy considers both affinity and priority of the \texttt{Bert-base} model. We prioritize inference of transformer models that operate on a single contextual text prompt over \acp{dnn} that operate on batches of input images. Upon on arrival of the inference request, affinity of \texttt{Bert-base} is towards GPU and priority of \texttt{Bert-base} is higher. Unlike Scenario-1, \texttt{Resnet-152} cannot be migrated to \ac{dla}, since \ac{dla} is occupied by \texttt{VGG-19}. Hence, our approach (i) freezes the execution of \texttt{ResNet-152} to accommodate \texttt{Bert-base} on GPU, (ii) \texttt{Bert-base} is mapped onto the GPU, (iii) \texttt{ResNet-152} is unfreezed and resumes execution on GPU once \texttt{Bert-base} finishes the execution at $t=0.71s$, (iv) Frequency of GPU is scaled up within the power constraints to address the performance loss of \texttt{ResNet-152} during task freezing. Our approach thus handles dynamic workload variation while minimizing the overall inference latency within the power constraints.  

\subsection{Related Work}\label{sec.related}

\noindent Widely used strategy for multi-\ac{dnn} inference on heterogeneous mobile platforms is to partition DNN into convolution blocks and execute them in a pipelined manner among different clusters~\cite{pipeit}. A similar approach has also been employed to pipeline transformer models~\cite{pipebert}, by splitting transformer encoder blocks across different clusters to maximize throughput. However, both these approaches~\cite{pipeit,pipebert} are confined to asymmetric CPUs and do not consider GPUs or DLAs. Edge AI inference techniques such as \textit{Omniboost}~\cite{omniboost} extend pipelining across \ac{cpu} and \ac{gpu} clusters, while \textit{MOC}~\cite{moc} proposes a multi-objective deep reinforcement learning agent that controls CPU cores and \ac{cpu}/\ac{gpu} frequencies. \textit{Tango}~\cite{tango} also uses a PPO-based RL agent for multi-DNN workloads to explore accuracy-performance-energy trade-offs to minimize inference latency. Moreover, \textit{Band}~\cite{band} demonstrated DNN partitioning across CPU, GPU, and NPU by generating subgraphs for DNNs. However, it only considers basic operators' support for DNNs, overlooking advanced operator support for transformers on DLA. Kim et al.~\cite{Interference_Aware} enhance multi-DNN workload scheduling by incorporating ML-based contention prediction, whereas \textit{RankMap}~\cite{rankmap} demonstrates a priority-aware multi-DNN manager that prevents DNN starvation under heavy execution loads. The aforementioned strategies primarily target CPU-GPU architectures, focusing on workload partitioning between different clusters. Advanced edge AI inference techniques have used domain-specific accelerators such as DLAs, TPUs, and NPUs. Kim et al.~\cite{ee_scenario_dl} explore \ac{cpu}, \ac{gpu}, and \ac{npu} resource allocation under varying latency constraints at runtime while minimizing energy consumption of the system. \textit{Axonn}~\cite{Axonn} focuses on distributing DNN layers based on energy and performance trade-offs across \ac{gpu}-\ac{dla} clusters. \textit{HaX-CoN}~\cite{HaX_CoNN} introduces shared memory contention-aware layer grouping and modeling inter-DSA layer transitions using a processor-centric slowdown model, to predict performance degradation. \textit{MapFormer}~\cite{mapformer} extends this by incorporating CPU, GPU, and DLA to support multi-DNN workloads, including transformer-based throughput and power estimation with \ac{dvfs}. Aforementioned multi-DNN scheduling strategies focus on: (i) unimodal DNN workloads for homogeneous tasks such as image classification and text classification, (ii) solutions tailored for fixed design-time systems, and (iii) do not consider run-time arrival of DNNs/Transformers/LLMs workloads. \textit{Twill} addresses these limitations through run-time adaptive scheduling of \ac{cai} systems with (i) affinity-aware cluster migration between \ac{gpu}/\ac{dla}, (ii) priority-aware task freezing, and (iii) adaptive \ac{dvfs} for power capping. 

\begin{table}
\vspace{-3pt}
\centering
\caption{Comparison of Runtime Techniques}
\vspace{-6pt}
\label{tab:runtime_comparison}
\scalebox{0.86}{
\renewcommand{\arraystretch}{1.2}
\begin{tabular}{|l|c|c|c|c|c|c|c|}
\hline
\textbf{Related Work} &~\cite{pipebert} &~\cite{moc} &~\cite{omniboost} &~\cite{tango} &~\cite{mapformer} &~\cite{HaX_CoNN} & \textbf{Twill} \\
\hline
\hline
Unknown app arrival & \xmark & \xmark & \xmark & \cmark & \xmark & \xmark & \cmark\\
\hline
Run-time exploration & \cmark & \cmark & \xmark & \cmark & \xmark & \xmark & \cmark\\
\hline
DLA clusters & \xmark & \xmark & \xmark & \xmark & \cmark & \cmark & \cmark\\
\hline
Task freezing & \xmark & \xmark & \xmark & \xmark & \xmark & \xmark & \cmark\\
\hline
Model partitioning & \xmark & \xmark & \xmark & \xmark & \cmark & \cmark & \cmark\\
\hline
Cluster migration & \xmark & \xmark & \xmark & \xmark & \xmark & \xmark & \cmark\\
\hline
DVFS tuning & \xmark & \cmark & \xmark & \cmark & \cmark & \cmark & \cmark \\
\hline
Encoder Transformer & \cmark & \xmark & \xmark & \xmark & \cmark & \xmark & \cmark \\
\hline
LLMs & \xmark & \xmark & \xmark & \xmark & \xmark & \xmark & \cmark \\
\hline
DNN workloads & \xmark & \cmark & \cmark & \cmark & \cmark & \cmark & \cmark\\
\hline
\end{tabular}
}
\vspace{-15pt}
\end{table}

\section{Twill Framework} \label{sec.arch}
\begin{figure*}
    \centering
    \vspace{-3pt}
    \includegraphics[width=0.95\linewidth]{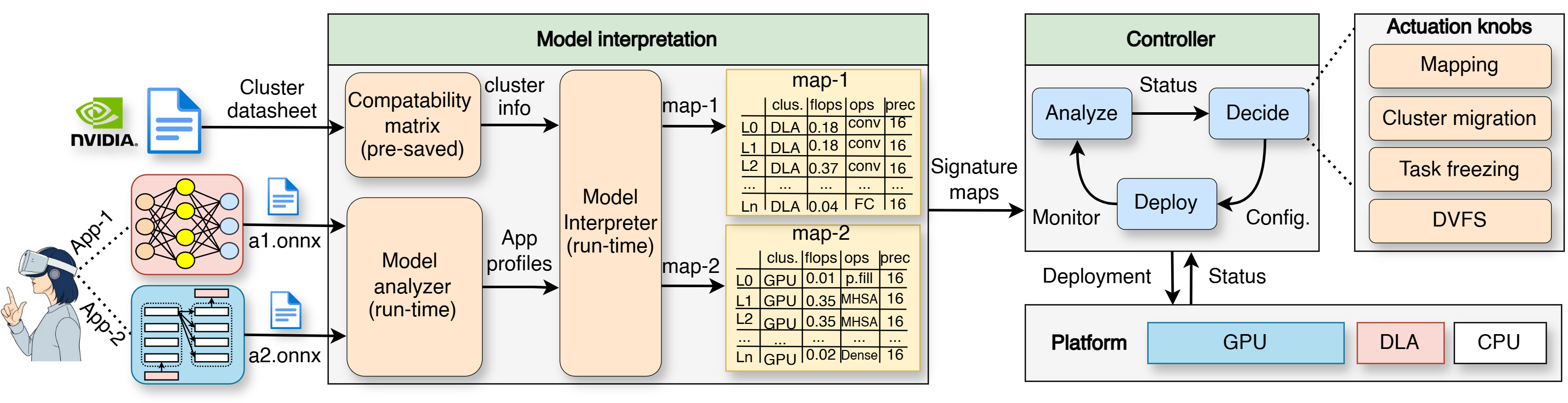}
    \vspace{-3pt}
    \caption{\textit{Twill} Framework including Model Interpreter to analyze the model-cluster affinity and Controller module performing run-time control knobs actuation.}
    \vspace{-12pt}
    \label{fig.system}
\end{figure*}
We have designed \textit{Twill} as an online scheduling framework to deploy \ac{cai} inference workloads on heterogeneous multi-core platforms hosting CPU, \ac{gpu}, and \ac{dla}. The platform receives run-time inference requests from the user, performs online heuristics exploration, and actuates the defined knobs to minimize the overall latency while guaranteeing a given power budget.
As shown in Figure~\ref{fig.system}, the \textit{Twill} framework includes two main modules, namely \textit{Model Interpreter} and \textit{Controller}. 
The \textit{Model Interpreter} performs online profiling to characterize each inference request. The interpreter includes a \textit{Model analyzer} to extract information from the model files, and a \textit{Compatibility matrix} to check the compatibility of the model layers with the available clusters. For each inference request, the \textit{Model Interpreter} extracts the characterization of the executed model and creates a \textit{Signature map} required for making actuation decisions. The \textit{Controller} monitors the \textit{Signature maps}, and the system status to make actuation decisions based on a set of defined actuation knobs and a heuristic algorithm. Then, the \textit{Controller} applies the workload and system configuration based on the actuation decision and continues monitoring the workload performance. Different modules of the \textit{Twill} framework are detailed in the following.


\subsection{Platform} 
We consider heterogeneous hardware platforms for edge/embedded computing, including multiple computing clusters; in particular, we focus on the Nvidia Jetson device family hosting CPU, \ac{gpu}, and \ac{dla}. 
Each cluster is provided with a \ac{dvfs} knob, with the only exception of the \ac{dla} in the platform considered in our experimental results, and the per-board power sensor.
The platform runs a standard \ac{os} as Linux, providing communication between the software modules to send and receive data and system commands. Specifically, the \ac{os} exposes interfaces for hardware control and application execution on various clusters. Platform-specific run-time frameworks are provided for executing applications on \ac{gpu} and \ac{dla}. 
We restrict the platform's power consumption to the \ac{tdp} limit, avoiding any physical damage to the board at a high temperature. 

\subsection{Workloads} 

\textit{Twill} targets the execution of \ac{cai} workloads representing a diverse set of \ac{dnn} and transformer models collaborating to compute a global output. 
These models execute different cognitive image and text processing tasks, including classification, detection, and generation. Moreover, the inference tasks may be (i) continuous inferences on streaming input data, such as sequences of images taken from a camera, (ii) single generation requests, e.g., of text outputs, or (iii) hybrid. Models in a single \ac{cai} workload may have data/precedence dependencies among each other that can be represented as a task graph; in fact, inference of some models is possibly triggered based on the outputs of a preceding model.
Even if the overall task graph of the \ac{cai} workload is known at design time, the actual execution presents several aspects that are dependent on the specific workload run and are unpredictable beforehand. In particular, the actual inference request to each model is specified at run-time based on the user requests or the results of the previous models in the task graph. This means that the execution time of each executed model is highly variable, for instance, based on the length of the input data stream to be processed by a \ac{dnn} or on the complexity of a generative request. As a consequence, the timing of all dependent models is affected. Therefore, the system experiences an unknown workload composed of multiple running models mixing \acp{dnn}, transformers, and \acp{llm}, each one arriving asynchronously. 



\subsection{Model Interpretation}
We design a model interpretation strategy as an online application profiling mechanism that creates a \textit{Signature map} of the given inference requests for the \textit{Controller} to schedule the \ac{cai} workload. 

\noindent\textbf{Model analyzer.} We use a \textit{Model analyzer} that parses the input \textit{ONNX} file describing the deep learning model using the \textit{ONNX-runtime} library~\cite{onnxruntime} to extract the layer information and metadata of the model. The \textit{Model Analyzer} also receives a user-defined \textit{application priority}, enabling the \textit{Controller} to select suitable candidates for making scheduling decisions while facing resource contention scenarios. \textit{Application priority} is a user-defined parameter that can be configured based on \ac{cai} workload requirements. For example, typical \ac{cai} systems prioritize user-driven prompt-based \acp{llm} and transformers over \acp{dnn} that run on streaming batches of inputs. We designed the \textit{Twill} framework to maneuver the workload scheduling decisions based on the changing application priorities. The \textit{Model analyzer} formulates an application profiling table (\texttt{App\_profile}) including the extracted model information and application priority. The \texttt{App\_profile} table includes elaborate model information, including model 1) name, 2) total parameters, 3) total layers, 4) total floating point operations, 5) layer types, 6) layer operation types, 7) layer-level input and 8) output sizes, and 9) activation functions.  

\begin{figure}[t]
    \centering
    \includegraphics[width=0.99\linewidth]{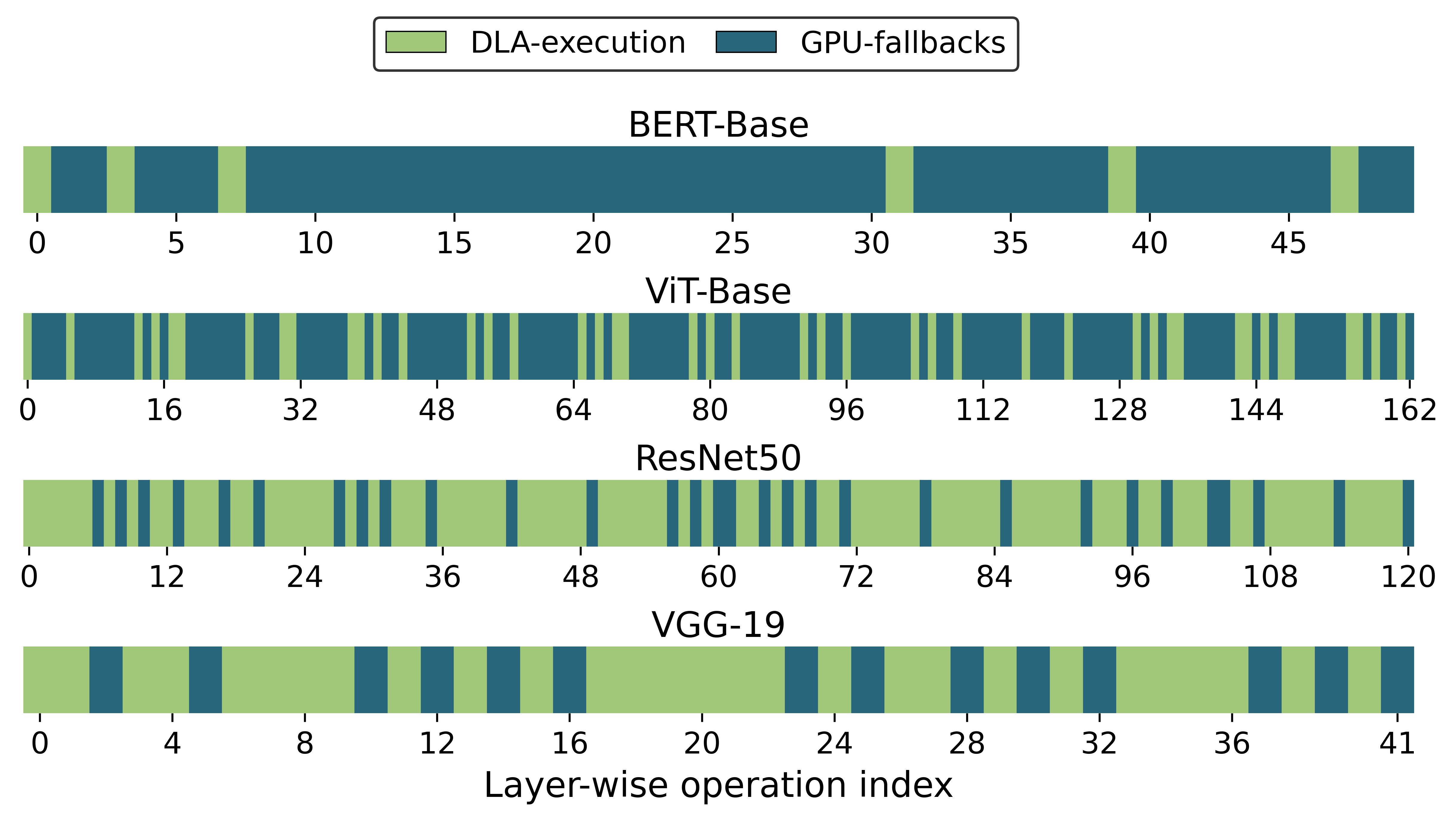}
    \vspace{-15pt}
    \caption{Supported operations on GPU and DLA clusters of Orin NX platform.}
    \vspace{-10pt}
    \label{fig.dla-ops}
\end{figure}


\noindent\textbf{Comparability Matrix.} 
We formulate a \ac{dla} compatibility matrix (\texttt{dla\_matrix}) from the official reference manual of Jetson Orin NX~\cite{dla_manual}. The matrix includes information about the supported and unsupported operations, precision level, layer types, spatial dimension, batch size, kernel size, and padding and stride ranges.
\textit{Twill} requires the cluster compatibility information because the available cluster can have dedicated support for a limited number of model operations. We demonstrate the compatibility of inferring \ac{dnn} and transformer models on the \ac{dla} cluster of Jetson Orin NX in Figure~\ref{fig.dla-ops}.
We separately ran inferences of 2 \ac{dnn} models (\texttt{ResNet50} and \texttt{VGG-19}) and 2 transformer models (\texttt{Bert-base} and \texttt{ViT-base}) on the \ac{dla} cluster. Figure \ref{fig.dla-ops} presents layer-wise operators supported by the \ac{dla} and \ac{gpu} fallback instances for unsupported operators. Most \ac{dnn} inference operations of \texttt{VGG-19} and \texttt{ResNet-50} are supported on \ac{dla}. For transformer models \texttt{Bert-base} and \texttt{ViT-base}, \ac{dla} supports only the pre-processing and convolution operations, while most of the operations fallback to the \ac{gpu} cluster. Therefore, the \ac{dla} cluster is more suitable to run \ac{dnn} workloads as compared to the transformers.

The \ac{dla} supports fixed-function layers required for \ac{dnn} inference, including convolution, pooling, activation, and fully connected layers. The \ac{dla} does not support the majority of computational operations in transformers, including multi-head attention, layer normalization, and advanced activation such as Gelu (Gaussian Error Linear Unit)~\cite{gelu}~\cite{rockchip_NPU, dla_manual}. 
For a model inference, \acp{dla} executes only the supported operators, and \textit{fallbacks} to the \ac{gpu} for running unsupported operations. 

\noindent\textbf{Model Interpreter.} The \textit{Model Interpreter} finds the model-cluster affinity based on the knowledge extracted from the application profile and the comparability matrix. 
The \textit{Model Interpreter} checks the layer-to-cluster affinity of each model through a mechanism shown in Algorithm~1. The algorithm takes as input the application signature (\texttt{App\_profile}) and DLA compatibility profile (\texttt{dla\_matrix}). At the start, the \textit{Model Interpreter} extracts the layer information from the application signature table (Line 2) and initializes a compatibility map (Line 3). The algorithm assigns GPU compatibility by default (Line 5) for each layer since GPUs can execute all operation types. The \texttt{DLACompatible} function (Lines 6--8) determines if a layer can also execute on DLA by checking three key conditions, (i) the model's precision must match DLA-supported precisions (Lines 11--13), (ii) the operation type must not be in the unsupported layer list (Lines 14--15), and (iii) for convolution and fully-connected layers, additional parameter constraints must be satisfied including kernel size, stride, and padding ranges (Lines 15-19). Finally, the \textit{Model Interpreter} creates a \textit{Signature map} for each model, including the per-layer cluster affinity, layer type, number of floating point operations, precision, and input-output shapes.

\begin{figure}[t]
    \centering
    \vspace{-6pt}
    \includegraphics[width=0.99\linewidth]{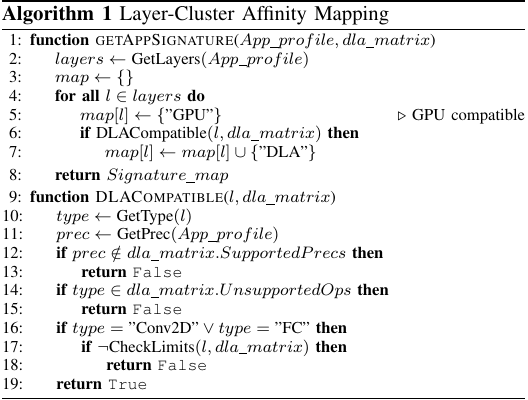}
    \vspace{-24pt}
\end{figure}



\begin{figure}[t]
    \centering
    \vspace{-6pt}
    \includegraphics[width=0.99\linewidth]{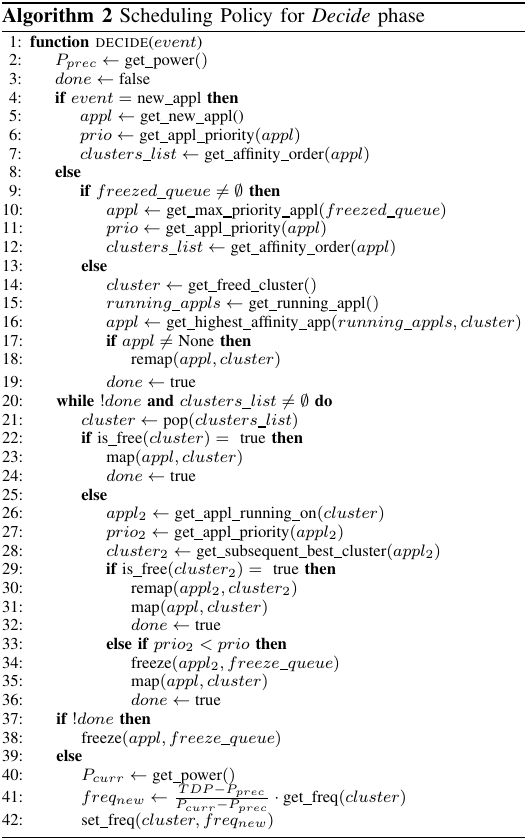}
    \vspace{-27pt}
\end{figure}

\subsection{Controller}
The \textit{Twill Controller} is the central scheduling mechanism that monitors the workload and system status, makes actuation decisions on various system-wide settings, and executes the model inferences. Conceptually, the controller is a middleware between the running workload and the hardware architecture. It accesses \ac{os} interfaces to monitor system status, in particular the power consumption and the current status of the various clusters. Then, it exploits the heartbeats sent by each running model to monitor their progress.

\noindent\textbf{Control knobs.} We defined an advanced set of actuation knobs to allow the controller to regulate workload deployment and execution; they are obtained by acting on the \ac{os} interface, opportunistically exploited to override the default \ac{os} scheduling and \ac{dvfs} governors. The knobs include \textit{Affinity-cluster mapping}, \textit{Cluster migration}, \textit{Task freezing}, and \textit{\ac{dvfs}}. 
\textit{Affinity-cluster mapping} is used to bind the inference request to a selected cluster; we map at most one model per cluster per time. In \textit{Cluster migration}, the controller splits the model layers and allocates the partitioned workload to a new cluster.
\textit{Task freezing} suspends the execution of a running application to free resources for another application. \ac{dvfs} enables frequency level setting on a selected cluster.

\noindent\textbf{Controller Policy.} We designed the \textit{Controller} as a feedback loop that runs in three steps, including \textit{Analyze, Decide}, and \textit{Deploy} phases at every control cycle. In the \textit{Analyze phase}, the \textit{Controller} waits for the events of new inference requests or release of a cluster due to termination or remapping of previously running inferences. The \textit{Controller} receives the \textit{Signature map} of each inference request from the \textit{Model Interpreter} and reads the current status of the available clusters. In the \textit{Decide} phase, the \textit{Controller} creates a mapping configuration for the given workload following a \textit{Scheduling policy}. Finally, in the \textit{Deploy} state, the workload is mapped to the clusters based on the configuration received from the \textit{Decide} state.

\noindent \textbf{Analyze.} In this phase, the controller waits for an inference request and transitions to the \textit{Decide} phase when a new request arrives or a previously running application completes execution. In case of a new inference request, the controller receives the \textit{Signature map} of each application from \textit{Model Interpreter}, and the system status, including cluster availability and \ac{gpu} frequency level. If a previously running application completes execution, the controller reads the system status and transitions to the \textit{Decide} phase to update the system configuration. The same event is notified in case an application is remapped in the previous control cycle. 

\noindent \textbf{Decide.} In this phase, the \textit{Controller} decides on the knobs actuation settings to achieve the minimum latency under a power budget. We consider \ac{tdp} as the power budget of the given platform. 
We have demonstrated the decision strategy of the \textit{Controller} in Algorithm~2. 
When a new application arrives, the \textit{Controller} extracts the application priority and the list of preferred clusters from the application signature maps acquired from the \textit{Model Interpreter} (Lines 4--7). Otherwise, if the event occurred due to a freed cluster, the \textit{Controller} first checks if any high-priority applications are waiting in the freeze queue and, if so, it gets the application data (Lines 9--12). If the queue is empty, the system evaluates whether an already-running application has higher affinity for the freed cluster (Lines 13--16); if so, such an application is remapped accordingly (Lines 17-19). Finally, if the algorithm entered the last branch since no application entered and the freeze queue is empty, the \textit{Decide} phase is completed.

After identifying the application and its preferred clusters, the \textit{Controller} iterates over the cluster list to attempt application-to-cluster mapping (Line 20--21). If a suitable cluster is available, the \textit{Controller} maps the application and exits the \textit{Decide} phase (Lines~22--24). If the preferred cluster is occupied, the \textit{Controller} checks if the currently-running application can be
remapped on a more suitable free cluster (Lines 26--27), and if this condition is true, the latter application is remapped and the newly arrived application is mapped on the freed cluster (Lines 29--32). 
Alternatively, the \textit{Controller} freezes the currently running lower priority application, and puts it in the freezing queue while making room for the new higher priority application (Lines 33--36).
Finally, if no suitable mapping has been found, the \textit{Controller} freezes the new application (Lines 37--38). 
After successful application-to-cluster mapping, the \textit{Controller} adjusts the frequency of the units provided with \ac{dvfs} (\ac{gpu} in our case), accordingly to maximize the power utilization while avoiding \ac{tdp} violations (Lines~39--42). In particular, power measures are taken before application start (i.e., at the previous control cycle, Line 2) and after the application start (Line 40); then, based to a linear model exploiting the available power budget and the measured power variation the new frequency is computed (Lines 41--42). Do note that in case we also have a remapped $appl_2$, the available power budget is split among the two handled models.

\noindent \textbf{Deploy.}
Finally, in the \textit{Deploy} phase, the \textit{Controller} deploys the scheduling configuration with knobs actuation decided in the \textit{Decide} phase. The \textit{Controller} enforces the configuration and transitions to the Analyze phase, waiting for a new event.  

\section{Experimental Evaluation} \label{sec.exp}
In this section, we present the experimental evaluation of our proposed \textit{Twill} framework in comparison with state-of-the-art edge AI inference strategies over relevant \ac{cai} workloads.


\subsection{Experimental Setup} 
\noindent\textbf{Controller prototype.} We have implemented the proposed \textit{Twill} framework in the Python programming language. The framework runs as a user-space process in a Linux \ac{os} environment. The controller monitors the run-time power using onboard power sensors.

\noindent\textbf{Platform and middleware.}
For evaluation, we use the Nvidia Jetson Orin NX~\cite{orin_nx} platform, which is composed of an Ampere GPU, an NVDLA (Nvidia's \ac{dla}), and hexa-core ARM A78 CPUs, along with 8 GB of RAM and GPU memory. The platform has onboard power sensors for collecting run-time power consumption of \ac{cpu}, \ac{gpu}, and \ac{dla} clusters. The platform hosts Linux-20.04 \ac{os} with CUDA support to enable GPU operations, and allows run-time actuation of GPU \ac{dvfs}. We enable the default \textit{Performance} scheduling governor for the GPU cluster to run our experiments. For our experimentation, we set \ac{tdp} as 10W \cite{orin_nx}. The average overhead of running the \textit{Twill} framework is 15ms on 100\% utilization of 1 \ac{cpu} core of the Orin NX platform.

\noindent\textbf{Workloads.}
For experiments, we considered AI models that can perform image classification, text classification, and text generation tasks, which can be chained to form widely used \ac{cai} systems. We use the LangChain~\cite{langchain} tool for orchestrating the \ac{cai} workload design. For vision applications, we consider \acp{dnn} -- \texttt{VGG-19}, \texttt{ResNet-50}, and \texttt{EfficientNet}-B4~\cite{vgg, resnet, efficientnet} and vision transformer models -- \texttt{ViT-base}, and \texttt{ViT-large}~\cite{vit}. For text classification, we consider encoder-based transformer models of \texttt{Bert-base} and \texttt{Bert}-large \cite{bert}. For \acp{llm}, we consider \texttt{DeepSeek R1}~\cite{deepseekr1} with 1.5 billion parameters and \texttt{Gemma 3} with 1 billion parameters. We implement the inference mechanism using the PyTorch~\cite{pytorch} framework and torchvision~\cite{torchvision}. For transformers, we use the Hugging Face~\cite{huggingface} library, and for \ac{llm} inference, we use Ollama~\cite{ollama}. 

\begin{figure*}
    \centering
    \vspace{-3pt}
    \includegraphics[width=0.95\linewidth]{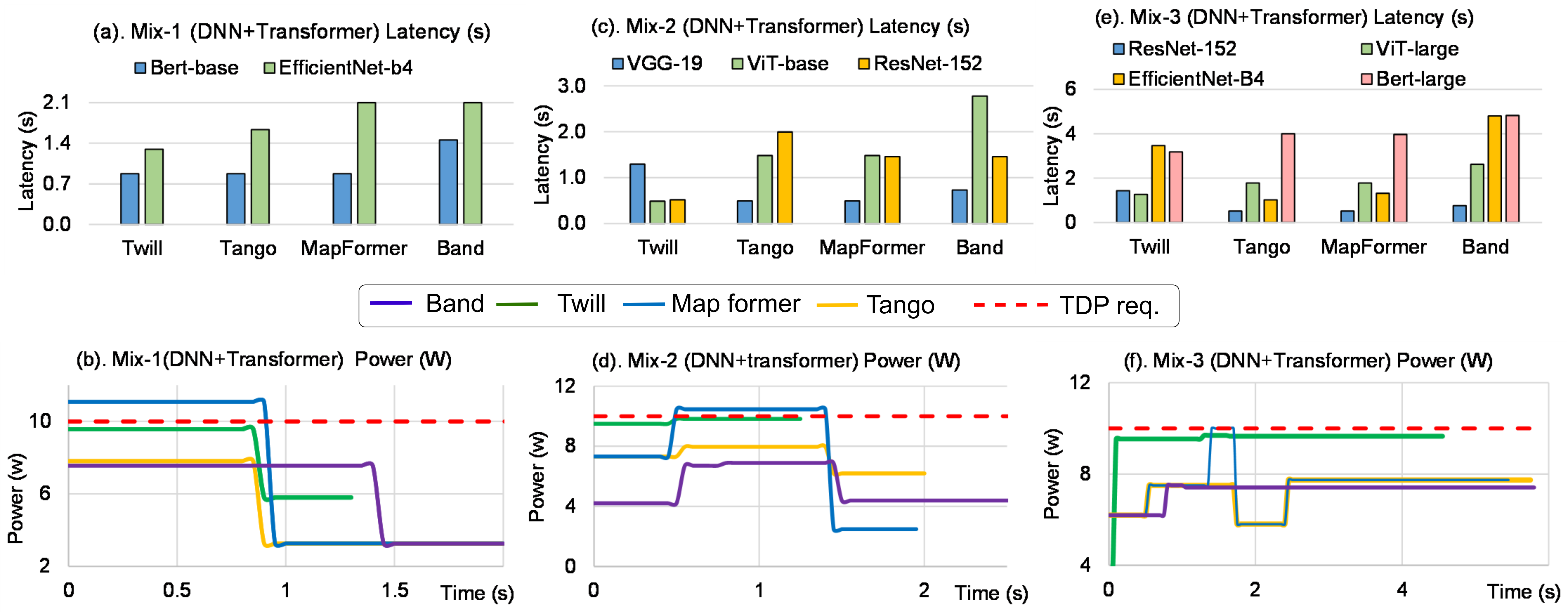}
    \vspace{-3pt}
    \caption{Inference latency and run-time power consumption of different workload mixes. The first row represents the application latency, and the second row presents the power consumption of each strategy.}
    \label{fig.mix-1-3}
    \vspace{-6pt}
\end{figure*}

\begin{figure}
    \centering
    \vspace{-3pt}
    \includegraphics[width=0.95\linewidth]{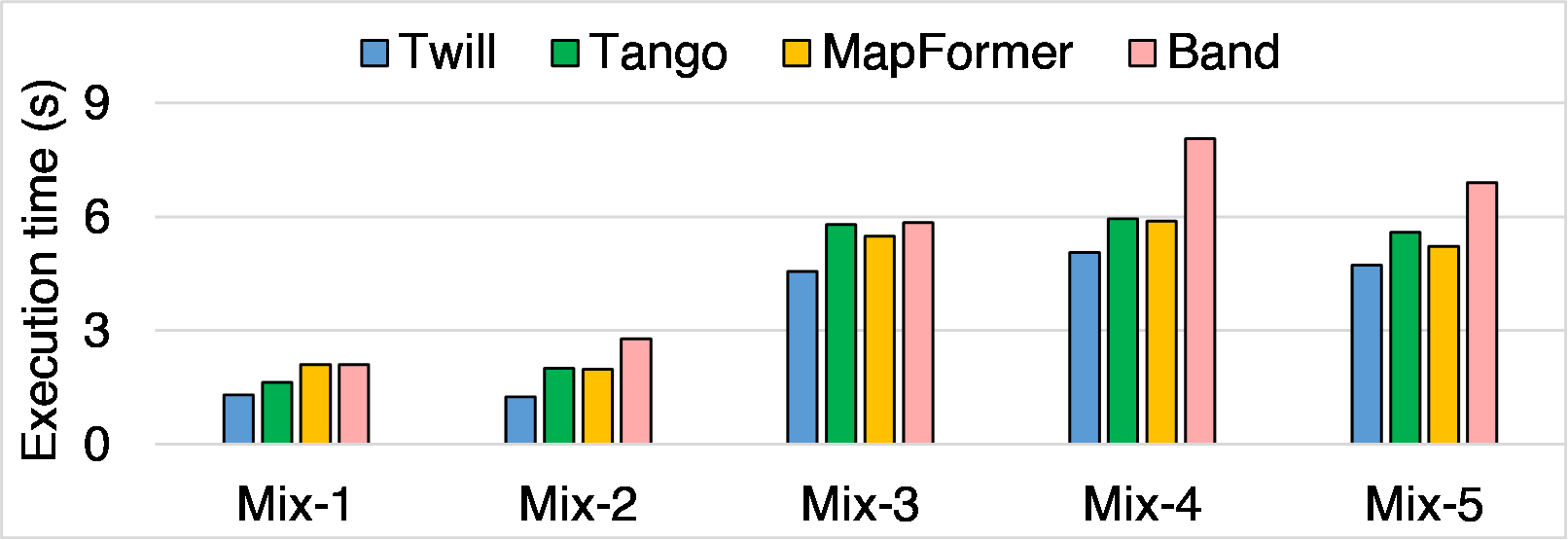}
    \vspace{-3pt}
    \caption{Overall \ac{cai} system execution time for mix 1-5.}
    \vspace{-18pt}
    \label{fig.mixes}
\end{figure}

\noindent\textbf{Comparison w.r.t. state-of-the-art approaches.} 
We consider three \ac{soa} edge AI inference strategies for heterogeneous platforms, including \textit{MapFormer} \cite{mapformer}, \textit{Tango} \cite{tango}, and \textit{Band} \cite{band}.
\textit{MapFormer} considers multi-DNN workloads for partitioning and allocating a suitable cluster between \ac{gpu} and \ac{dla} while scaling the \ac{dvfs} to manage power and throughput. This strategy makes design-time mapping decisions, assuming all inference requests arrive simultaneously. We implemented a transformer-based estimator of \textit{MapFormer} that predicts throughput and power consumption distributions for multi-DNN workloads. \textit{Tango} is a run-time multi-DNN workload management strategy on \ac{cpu} and \ac{gpu} clusters using reinforcement learning based exploration. We used Gymnasium library \cite{gym} to implement \textit{Tango's} latency estimation model. \textit{Band} partitions the multi-DNN workload into subgraphs at design-time, and maps the subgraphs based on the supported \ac{dnn} operations on GPU and NPU clusters. We implemented the \textit{Band's} model analyzer to generate the subgraphs for cluster mapping based on the \ac{dnn} operations and FLOP counts.

\noindent\textbf{Evaluation metrics.} We measure per-application inference latency, throughput of the entire workload mix, and the platform's power consumption for our experimentation. For \acp{dnn}, we measure the latency as the time required to infer a batch of input images. For transformers, we measure the latency as the inference of a text prompt for a classification task. For \acp{llm}, the inference latency varies depending on the number of generated output tokens against a text prompt. 
We measure throughput as the number of inferences per second, reflecting the system's capacity to handle dynamic \ac{cai} workloads.

\subsection{Experimental Results}
\begin{figure}
    \centering
    \vspace{-3pt}
    \includegraphics[width=0.95\linewidth]{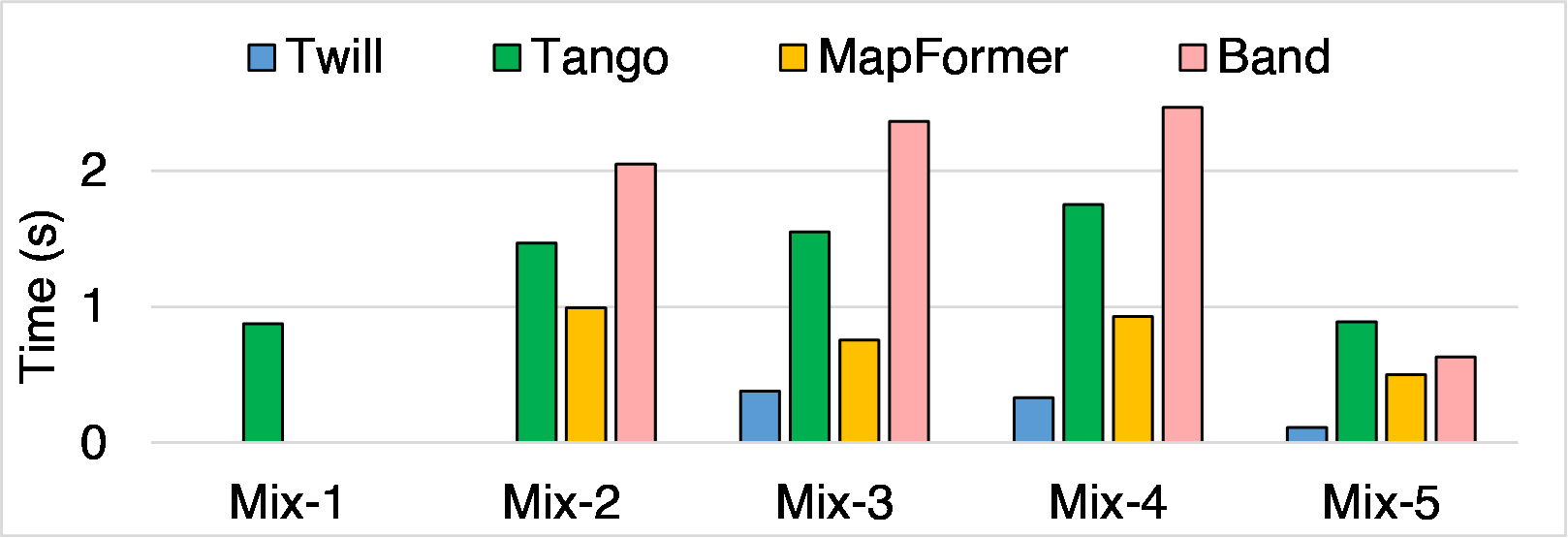}
    \vspace{-3pt}
    \caption{Waiting time of applications in the execution queue across mixes.}
    \vspace{-18pt}
    \label{fig.wait-time}
\end{figure}

For evaluation, we create different \ac{cai} workload combinations (\textit{Mix
 1-5}), encompassing various degrees of composability using multiple \acp{dnn}, transformers, and \acp{llm}. Workload \textit{Mixes} 1-3 are based on \ac{dnn} and transformer models, and the \textit{Mixes 4-5} are based on \acp{dnn}, transformers, and \ac{llm}.
Figure~\ref{fig.mixes} shows the execution time of different \ac{cai} workload mixes with \textit{Twill} and three SoA edge AI inference strategies. Our proposed \textit{Twill} strategy has the lowest execution time in comparison with other relevant strategies, achieving up to 38\%, 54\%, 22\%, 37\%, and 31\% lower execution time for \textit{Mix-1} to \textit{Mix-5}, respectively. \textit{Twill} jointly tunes multiple actuation knobs of affinity-cluster mapping, task freezing, cluster migration, and \ac{dvfs} to minimize the inference latency within the available power budget. This coordinated joint actuation for run-time varying \ac{cai} workloads results in significantly lower execution time. Other strategies are confined to offline analysis of workload characteristics and can only handle inference request arrivals that are known at design time. Hence, they minimize the inference latency of the first and/or second arriving applications, while resulting in significantly higher overall inference latency as the number of concurrent inference requests and compute diversity increases. On average, \textit{Twill} has 20\%, 19\%, and 34\% lower execution time than \textit{Tango}, \textit{MapFormer}, and \textit{Band}, respectively. Figure~\ref{fig.wait-time} shows the time each inference request spent in the waiting queue due to shared resource contention. \textit{Twill} successfully avoids inference request queuing in \textit{Mix-1} and \textit{Mix-2}. For \textit{Mixes} 3-5, \textit{Twill} reports 83\%, 87\%, and 88\% lesser waiting time on average than other relevant strategies. Evaluation metrics of each workload mix are detailed in the following. 



\noindent\textbf{Workload Mix-1.} \textit{Mix-1} includes inference requests of \texttt{Bert-base} and \texttt{EfficientNet-b4} models with \texttt{EfficientNet-b4} arriving $20ms$ after \texttt{Bert-base}. This creates a scenario of progressively increasing workload with both \ac{dnn} and transformer applications, such that both applications run concurrently at $t=20ms$. Figure~\ref{fig.mix-1-3}(a) shows the latency of both \texttt{Bert-base} and \texttt{EfficientNet-b4} against different strategies.
\textit{Tango} does not support \ac{dla} execution, and puts \texttt{EfficientNet-b4} in the waiting queue while executing \texttt{Bert-base} on \ac{gpu} causing a waiting overhead for the \ac{dnn} model.
Similarly, \textit{Mapformer} first allocates \ac{gpu} to \texttt{Bert-base}, and \ac{dla} to \texttt{EfficientNet} once it arrives. This strategy decides the workload allocation at an application arrival and does not support run-time configuration changes for applications with unknown arrivals. Hence, \texttt{EfficientNet} continues executing on a lower performing \ac{dla} even when \ac{gpu} becomes available after the execution \texttt{Bert-base}. Finally, \textit{Band} also allocates \ac{gpu} to \textit{Bert-base}, and \ac{dla} to EfficientNet, without \ac{dvfs} control. The strategy depends on the default Linux governors for \ac{dvfs} control and is constrained to making workload partitioning decisions upon arrival. Finally, \textit{Twill} maps \texttt{EfficientNet} on \ac{dla} until \texttt{Bert-base} executes on the \ac{gpu} cluster. \textit{Twill} migrates \texttt{EfficientNet-b4} to \ac{gpu} for faster execution once \texttt{Bert-base} terminates, at $t=874ms$. Moreover, \textit{Twill} scales the \ac{gpu} frequency following the given workload to maximize the power budget utilization for better performance. Hence, \textit{Twill} capitalizes on run-time knob actuation decisions for unknown workloads scenarios, ensuring faster workload execution while staying within the available power budget.  Figure~\ref{fig.mix-1-3}(b) shows the power consumption of each strategy during the workload execution. \textit{Tango} has lesser power consumption because it runs one application on \ac{gpu} for any given instance and does not involve \ac{dla}. \textit{Band} does not actively actuates the \ac{dvfs} which reduces the overall power consumption. \textit{Mapformer} actively tunes \ac{dvfs} shows \ac{tdp} violations for 48\%  of the entire execution time because it does not cater to the power budget. \textit{Twill} ensures the maximum utilization of the available power budget to achieve minimum latency.

\noindent\textbf{Workload Mix-2.} For workload \textit{Mix-2}, we introduce three inference applications \texttt{VGG-19}, \texttt{ViT-base}, and \texttt{ResNet-152}, representing a different scenario from \textit{Mix-1} in terms of dynamicity and computational load (Figure~\ref{fig.mix-1-3}(c)). \texttt{VGG-19} arrives first and is mapped to the \ac{gpu} by \textit{Tango}. At $t=20ms$, \texttt{ViT-base} arrives with high GPU affinity, but \textit{Tango} cannot migrate \texttt{VGG-19} to \ac{dla}, resulting in high wait time. After \texttt{VGG-19} finishes at $t=494ms$, \texttt{ViT-base} runs on the \ac{gpu} until $t=1484ms$. Meanwhile, \texttt{ResNet-152} arrives at $t=1000ms$ and waits until the GPU is free.
\textit{Mapformer} maps \texttt{VGG-19} to GPU and queues \texttt{ViT-base}, later assigning it to GPU and \texttt{ResNet-152} to \ac{dla}. \textit{Band} follows a similar mapping but suffers from higher latency due to lacking \ac{dvfs} control. \textit{Twill} minimizes total execution time by initially assigning \texttt{VGG-19} and \texttt{ViT-base} to \ac{dla}, scaling GPU frequency, and mapping \texttt{ResNet-152} to GPU. Once \texttt{ResNet-152} completes, \texttt{VGG-19} is migrated to GPU. Figure~\ref{fig.mix-1-3}(d) shows runtime power usage, where \textit{Mapformer} exceeds the \ac{tdp} limit for 49\% of the execution time.

\begin{figure}
    \centering
    \vspace{-3pt}
    \includegraphics[width=0.95\linewidth]{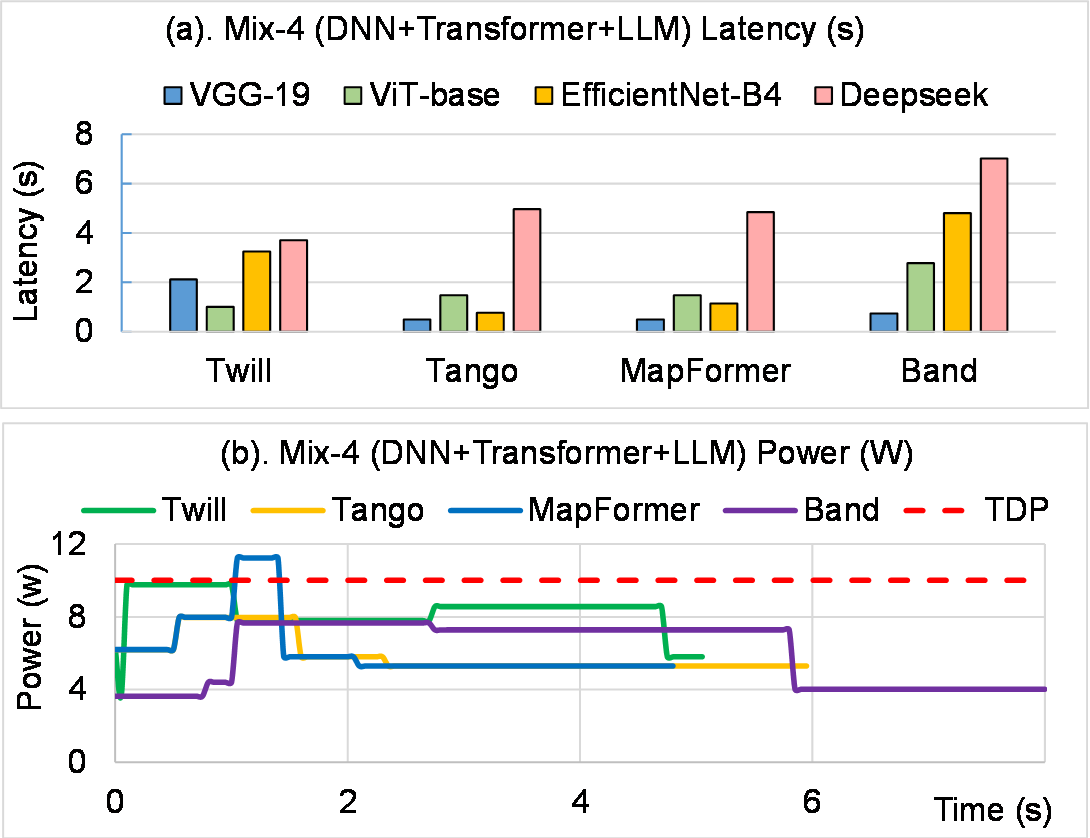}
    \vspace{-3pt}
    \caption{Inference latency and run-time power consumption of \textit{Mix-4}}
    \vspace{-18pt}
    \label{fig.mix-4}
\end{figure}

\noindent\textbf{Workload Mix-3.}
In this workload mix shown in  Figure~\ref{fig.mix-1-3}(e), we progressively introduced four applications to the system, including two \acp{dnn} (\texttt{ResNet-152} and \texttt{EfficientNet-b4}), and two transformer models (\textit{ViT-large}, and \texttt{Bert-large}). At t=0, \texttt{ResNet-152} arrives and \textit{Tango} maps it to the \ac{gpu} cluster, while \texttt{ViT-base} waits in a queue after arriving at $t=20ms$. When \texttt{ViT-base} is executing, \texttt{EfficientNet-b4}, \texttt{Bert-large} arrive at $t = 1000 ms$ and $t = 10020 ms$, respectively, and wait in the queue for later execution. \textit{MapFormer} executes \texttt{ResNet-152}, and puts \texttt{Vit-large} in waiting queue. Later, it maps \texttt{ResNet} on \ac{dla}, while \texttt{Bert-large} again waits in the queue for termination of \texttt{ViT-large}. \textit{Band} again has similar mapping as \textit{Mapformer} without \ac{dvfs} scaling. \textit{Twill} successfully runs \texttt{ResNet-152}, and \texttt{ViT-large} in parallel by mapping them on \ac{dla}, and \ac{gpu} successfully. Later, it maps \texttt{Bert-base} on \ac{gpu}, while freezing \texttt{ResNet} because \texttt{Bert-base} has higher \ac{gpu} affinity. Finally, \textit{ResNet} first runs on \ac{dla} after \texttt{VGG-19} finishes execution and migrates to \ac{gpu} once \texttt{Bert-base} finishes execution. Figure~\ref{fig.mix-1-3}(f) shows the power consumption during the workload execution. \textit{Mapformer} shows TDP violations for 7\%  of the entire execution.

\noindent\textbf{Workload Mix-4.}
In \textit{Mix-4}, we considered a highly heterogeneous workload considering two \acp{dnn} (\texttt{VGG-1}, \texttt{Efficientnet-b4}), one encoder transformer (\texttt{ViT-base}), and an \ac{llm} (Deepseek-R1) as shown in Figure~\ref{fig.mix-4}(a). The generative model has a different number of output tokens depending on the input prompt, and for a quantifiable analysis of the output latency, we considered the latency of 100 output tokens. \textit{VGG-19} arrives at $t=0$ and \textit{Tango} maps the application to \ac{gpu}, while \texttt{ViT-base}, arriving at $t=20ms$, has to wait in the queue for the prior application to finish its execution. Similarly, \texttt{EfficientNet} arriving at $t=1000ms$, and \textit{Deepseek} arriving at $t=1020ms$, have to wait in a queue for their previous application to finish execution. This long waiting overhead for each application slows down the workload execution. Similarly, \textit{Mapformer} and \textit{Band} also place \texttt{ViT-base} in the waiting queue while executing \texttt{VGG-19}. Later, these strategies map \texttt{EfficienNet} on \ac{dla}, while \texttt{Deepseek} starts execution after \texttt{ViT-base} terminates on \ac{gpu}. \textit{Twill} maps \texttt{VGG-19} on \ac{dla}, and \texttt{ViT-base} on \ac{gpu}. Later, \texttt{DeepSeek} is mapped on \ac{gpu} having higher affinity to \ac{gpu} cluster. \texttt{Efficientnet} has to wait for \texttt{VGG-19} for $110ms$ to terminate on \ac{dla}. Later \texttt{EfficientNet} runs on \ac{dla} until \texttt{Deepseek} finishes execution on \ac{gpu}, and is migrated to \ac{gpu} until execution. Figure~\ref{fig.mix-4}(b) shows that only \textit{Mapformer} shows TDP violations for 7\%  of the entire execution time, while other strategies stay within the available budget.

\begin{figure}
    \centering
    \vspace{-3pt}
    \includegraphics[width=0.95\linewidth]{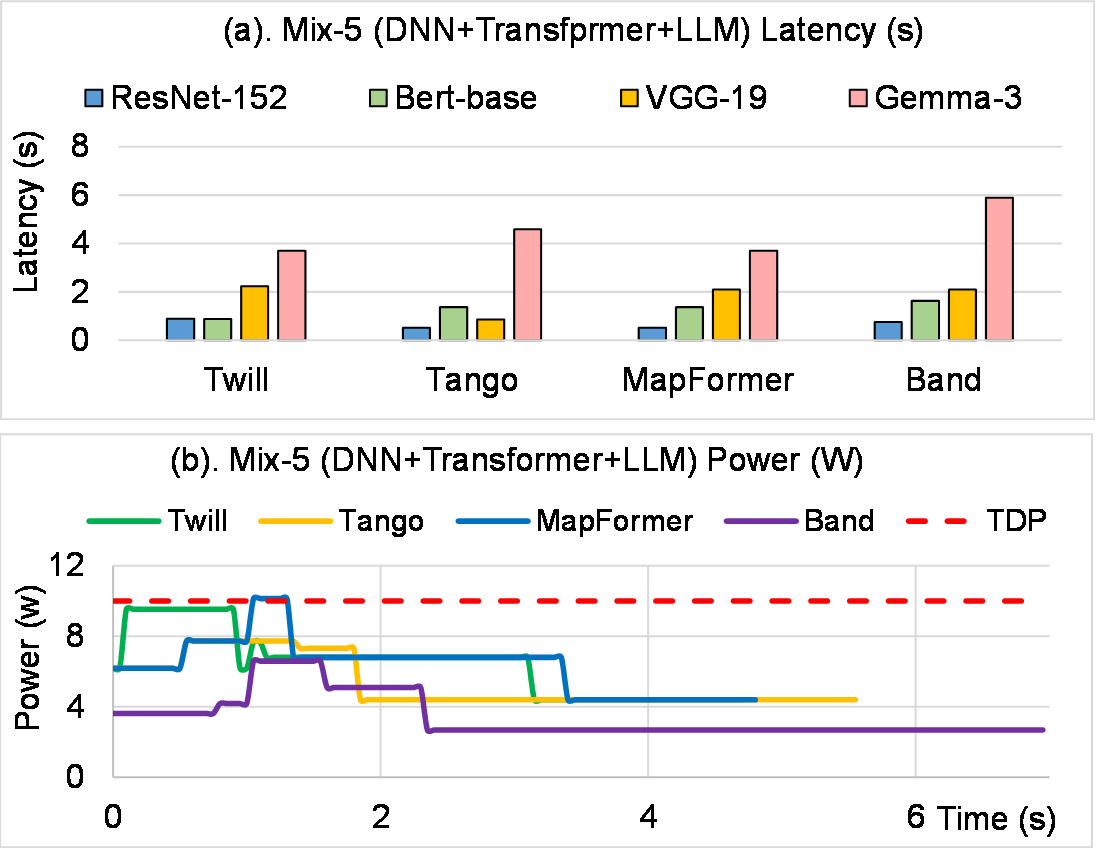}
    \vspace{-3pt}
    \caption{Inference latency and run-time power consumption of \textit{Mix-5}}
    \vspace{-15pt}
    \label{fig.mix-5}
\end{figure}

\noindent\textbf{Workload Mix-5.} 
In the final workload \textit{Mix-5} shown in Figure~\ref{fig.mix-5}(a), we considered four workload application including \texttt{ResNet-152}, \texttt{Bert-base}, \texttt{VGG-19}, and \texttt{Gemma-3} introduced at $t=0ms$, $t=20ms$, $t=1000ms$, and $t=1020ms$ progressively to the system. \textit{Tango}, maps each application on \ac{gpu}, where each application other than the first \textit{ResNet-152} has to wait in the queue for execution. \textit{Mapformer}, and \textit{Band} map \texttt{ResNet-152}, \texttt{Bert-base}, and \texttt{Gemma} on \ac{gpu} and \texttt{VGG-19} on \ac{dla} where \texttt{Bert-base} waits in the queue while \ac{gpu} is occupied by \texttt{ResNet}. \textit{Twill} intelligently migrates \texttt{ResNet} between \ac{dla} and \ac{gpu} while making room for \texttt{Bert-base}, and \texttt{Gemma} on the \ac{gpu} cluster. \texttt{VGG-19} is mapped on \ac{dla} for parallel execution with \texttt{Gemma}.  Figure~\ref{fig.mix-5}(b) reports 6\% \ac{tdp} violations of \textit{Mapformer}, while the other strategies remain within the available power budget.
\vspace{-3pt}

\section{Conclusions}\label{sec.conc}
We presented \textit{Twill}, an adaptive run-time framework for scheduling \ac{cai} workloads on heterogeneous mobile edge platforms. Our proposed approach uses affinity-aware mapping and migration, priority-aware task freezing/unfreezing, and \ac{dvfs} to handle concurrent inference requests, while minimizing inference latency within power budgets. Experimental evaluation of our strategy over contemporary \ac{cai} workloads against relevant edge AI inference techniques demonstrated up to 54\% lower inference latency while honoring power budgets. Orchestrating multi-LLM \ac{cai} systems on mobile platforms is planned as our future work.

\section*{Acknowledgments}
This work is funded by the European Union's Horizon 2020 Research and Innovation Program (APROPOS) under the Marie Curie grant No. $956090$

\bibliographystyle{IEEEtran}
\bibliography{references}

\end{document}